**Rhythm-consistent semi-Markov simulation of tourist mobility rhythms with probabilistic event-to-POI assignment: Hakone, Japan**

Jianhao Shi[1], Tomio Miwa[1*], Wanglin Yan[2]
[1]Nagoya University, [2]Keio University
[*]Corresponding author: miwa@nagoya-u.jp

**Abstract**
Understanding the timing and sequencing of activity participation in tourist mobility is central to travel behavior research, yet GPS trajectories are noisy, irregularly sampled, and only weakly linked to activity locations, which limits interpretation and scenario analysis. We address this by mapping each stay event to candidate points of interest (POIs) probabilistically, using explicit prior–likelihood weighting that yields a normalized compatibility distribution rather than hard matching. Using one month of high-density tourist trajectories in Hakone, Japan (November 2021), we construct semantic stay-event sequences based on observed place-category labels (MID10) and describe mobility rhythms through hour-by-category profiles, category transitions, and expected dwell patterns. Building on these rhythm signatures, we develop a rhythm-consistent semi-Markov simulator that generates synthetic stay-event sequences with time-conditioned transitions and category-dependent dwell behavior. In the observed data, hour-by-category summaries are computed by probability-weighted aggregation over soft labels; in simulation, each event is generated with a discrete category and a sampled dwell duration, enabling like-for-like comparison after aggregation. We further conduct counterfactual POI-inventory scenarios to quantify how hypothetical POI configuration changes shift stay intensity across time, categories, and space, particularly around hubs and main corridors. Observed–simulated comparisons show close agreement in temporal profiles and category distributions, indicating that probabilistic labeling and rhythm-consistent simulation preserve key mobility structure while providing an interpretable basis for transport-geography scenario evaluation.

**1. Introduction**
Large-scale positioning datasets are increasingly used in travel behaviour research to examine when and where people engage in activities and how those patterns vary across space and time. However, in tourism contexts, irregular sampling, measurement noise, and weak (often ambiguous) links between observed stays and points of interest (POI) semantics can limit behavioral interpretation and the design of place-based interventions (Song et al., 2019; Kato et al., 2024; Alessandretti et al., 2020). In this study, POI semantics denotes the place-category semantics (MID10 POI categories) used to interpret stays in terms of the functional meaning of visited places; it does not assume an externally defined activity ontology beyond the observed place category.

A practical response is to shift the analytical unit from raw polylines to event sequences of stops/stays with explicit dwell and temporal ordering. This event-based framing is aligned with classical tourist movement modeling, wherein transitions and durations are jointly modeled (Xia et al., 2011). However, when trajectories are reconstructed from sparse observations or aligned to networks, any uncertainty that is introduced upstream can propagate to event detection and dwell estimation (Forghani et al., 2020; Dalla Torre et al., 2019), which, in turn, intensifies the semantic ambiguity at the event-to-place linkage stage, especially when multiple plausible POIs exist near a single stay centroid.

POI-aware mobility analytics has advanced from urban-function measures and facility interaction patterns (Yue et al., 2017; Yu et al., 2017) to structured semantic representations such as city semantic diagrams and graph-based activity inference (Shan et al., 2022; Liu, Wu, Peng, et al., 2022), as well as optimization-based visited-POI assignment that enforces sequence consistency (Suzuki et al., 2019). Nevertheless, many pipelines still collapse event–place linkage into a single best match, which becomes brittle in dense or multifunctional POI environments—exactly where calibrated uncertainty is required for robust planning interpretation. This issue is particularly salient when an event is interpreted as belonging to a category such as SpaOnsen: multiple candidate facilities typically exist, and the planning-relevant question is not only "which category" but also "which destination among alternatives" should be modeled explicitly, rather than implicitly fixed by nearest-neighbor assignment.

Scenario readiness further tightens the requirement. Planning-oriented "what-if" questions under POI changes (type conversion, additions, relocations, and removals) require models that maintain controllable micro-level event sequences and simultaneously preserve urban rhythm and semantic compositions. In this study, urban rhythm refers to a reproducible time-of-day structure in stay intensity and category composition (e.g., hour×category profiles and transition tendencies) that is empirically observable and operationally meaningful for destination management. Existing work provides the building blocks via POI embeddings and demand modeling (Jiang et al., 2021; Liu et al., 2017) and synthetic mobility generation under limited labels or privacy constraints (Cui et al., 2021; Smolak et al., 2020); however, it does not typically ensure that simulated sequences are simultaneously rhythm-consistent and uncertainty-aware at the event–POI semantic layer.

We propose a probabilistic framework for event-based semantic mobility modeling and rhythm-consistent simulation by (i) converting Global Position System (GPS) trajectories into ordered stay events; (ii) performing a likelihood-based probabilistic event-to-POI assignment using an explicit prior-likelihood weighting, which yields a posterior-like score over candidate POIs rather than hard matching; and (iii) simulating synthetic stay-event sequences that preserve observed mobility rhythm and category-level semantic structure. The latter two stages use different but compatible representations: in the real data, hour×category summaries are computed by expected (probability-weighted) aggregation; in the simulation, each stay event is generated with a discrete category and a dwell duration sampled from the learned temporal-category structure, and destination choice, within a selected category, is determined by an explicit spatial-selection mechanism over candidate POIs. This pipeline directly supports travel behavior inference by enabling scenario-based evaluation of POI interventions—for example, which POI configurations are likely to redistribute stay intensity by hour and category, and how such shifts propagate along event sequences—under a controlled and interpretable behavioral mechanism (Park et al., 2023; Beritelli et al., 2020).

The remaining paper is organized as follows. Section 2 reviews related work along the pipeline from trajectory reconstruction and event extraction to semantic enrichment, sequence modeling, and simulation. Section 3 introduces the proposed methodology. Sections 4–5 present empirical results and scenario-based analyses, followed by the discussion and conclusions.

**2. Literature Review**
2.1 Mobility data sources and trajectory reconstruction
Mobility studies draw on heterogeneous positioning data, from high-frequency GPS to sparse cellular observations. Cellular-based mobility requires careful reconstruction before it can be subject to behavioral modeling, and sparse settings motivate specialized map-matching strategies (Forghani et al., 2020; Dalla Torre et al., 2019). Although foundational, these approaches inherit a persistent limitation: reconstruction uncertainty is rarely propagated explicitly to downstream steps (e.g., stop detection, dwell estimation, and POI linkage), despite directly affecting semantic interpretability. Privacy and data-sharing constraints further motivate synthetic mobility generation as an enabling layer. Modular simulation frameworks reproduce spatiotemporal statistics while reducing disclosure risk (Smolak et al., 2020), positioning simulation as both a methodological tool and a practical instrument for scenario experimentation when raw trajectories cannot be shared.

2.2 From trajectories to events: segmentation, stops, and dwell extraction
Event extraction is decisive under irregular sampling. Space–time interpolation, combined with density-based clustering, supports segmentation when the logging frequency is inconsistent (Song et al., 2019). Indeed, stay-point detection outcomes can vary substantially with GPS log intervals, implying that algorithm and parameter choices must be justified relative to data resolution (Kato et al., 2024). Trip-end identification, via spatiotemporal clustering, complements this perspective by distinguishing meaningful stops from transient congestion or redundant pauses (Yao et al., 2022). In high-precision environments, real-time stop monitoring using ultra-wideband streams illustrates the operational value of event representations for visitor management (Hachem et al., 2024). Collectively, these studies justify event-centric modeling but remain open-ended on the question of representing semantic ambiguity when events are linked to dense POI sets.

2.3 Semantic enrichment with POIs and urban meaning
POI data can be used to measure urban function and activity intensity, including POI-based mixed use and its association with vibrancy (Yue et al., 2017) and POI co-location patterns with network neighborhoods and distance-decay effects (Yu et al., 2017). Recent CEUS-oriented work further explores POI networks for functional organization (Lin et al., 2025). At the individual level, semantic diagrams and graph representations couple trajectories with POIs to infer activities (Shan et al., 2022; Liu, Wu, Peng, et al., 2022), while optimization-based visited-POI assignment enforces sequence-level consistency beyond local nearest-neighbor matching (Suzuki et al., 2019). A recurring limitation in planning-oriented pipelines is the use of a single semantic label per stop. Under POI interventions, hard labeling can systematically bias scenario inference by over-committing ambiguous events to one category and distorting downstream hour×category rhythm comparisons.

2.4 Sequence and transition modeling: from probabilistic baselines to deep architectures
For event sequences, classical semi-Markov models capture both transitions and durations (Xia et al., 2011), making them valuable for tourism research. Markov-chain mixture models further characterize population heterogeneity in activity patterns (Zhou et al., 2021). Probabilistic baselines for next-place/POI prediction include hidden Markov model-based mobility prediction at POIs (Lv et al., 2017) and variational embeddings of hidden Markov models that connect structured latent dynamics to richer representations (Jeong et al., 2021). Deep models incorporate spatiotemporal constraints via gated architectures (Zhao et al., 2022) and hierarchical attention mechanisms for next-POI tasks (Liu, Guo, & Qiao, 2022). Beyond individual sequences, community discovery from mobile phone data reveals stable spatial interaction structures (Gao et al., 2013), and dynamic tensor factorization provides a general approach to pattern discovery in spatiotemporal systems (Chen et al., 2025).

While the aforementioned studies span a broad methodological spectrum, they are primarily designed for prediction or pattern discovery rather than for scenario-ready planning evaluation under POI configuration changes. In particular, they generally do not couple micro-level urban rhythm preservation with calibrated semantic uncertainty at the event–POI layer, which is critical when simulated sequences are used to support counterfactual "what-if" assessment for planning applications.

2.5 Simulation and scenario analysis: from synthetic schedules to flow generators
Simulation links behavioral modeling to scenario evaluation. Under limited labels and low-frequency GPS, probabilistic modeling can generate daily activity–location schedules that match aggregate distributions (Cui et al., 2021). Privacy-preserving simulation similarly emphasizes the reproduction of statistics under shareability constraints (Smolak et al., 2020). At macro scales, deep gravity models generate mobility flows while exposing feature relevance

(Simini et al., 2021), multi-scale models unify mobility across urban agglomerations (Chen et al., 2023), and mobility-scale theory clarifies how modeling granularity relates to meaningful displacement and dwell scales (Alessandretti et al., 2020).

For planning, activity-based simulation can evaluate interventions with explicit equity considerations (Somanath et al., 2025). In tourism and destination design, corridor-based and destination-level analyses connect mobility patterns to planning decisions (Park et al., 2023; Beritelli et al., 2020), and POI demand modeling links mobility patterns to facility-level impacts (Liu et al., 2017). One need remains: an integrated micro-level simulator that preserves empirical urban rhythm while treating event–POI semantics probabilistically, because, otherwise, hard-labeled semantics can translate POI supply perturbations into biased scenario conclusions (e.g., inflated or suppressed category-level shifts) even when the underlying behavioral mechanism is unchanged.

2.6 Summary and positioning of this study

Prior work establishes strong foundations for (i) trajectory reconstruction under imperfect positioning data, (ii) event extraction under irregular sampling, (iii) POI-based semantic enrichment, (iv) sequence modeling with temporal structure and heterogeneity, and (v) synthetic mobility generation for privacy or scenario exploration. For destination planning, an underdeveloped intersection is a pipeline that is simultaneously event-based, uncertainty-aware in event–POI semantics, rhythm-consistent in terms of urban rhythm, and scenario-ready under POI configuration changes (Park et al., 2023; Beritelli et al., 2020).

**3. Methodology**

3.1 Study area and data sources

This study focuses on the Hakone area in Kanagawa Prefecture, Japan. The analysis domain is defined by a fixed study boundary that is used consistently for (i) filtering stay events from raw mobility traces and (ii) curating a POI inventory used in both baseline and counterfactual scenario simulations. Fig. 1 summarizes the study area and the spatial distribution of POI categories.

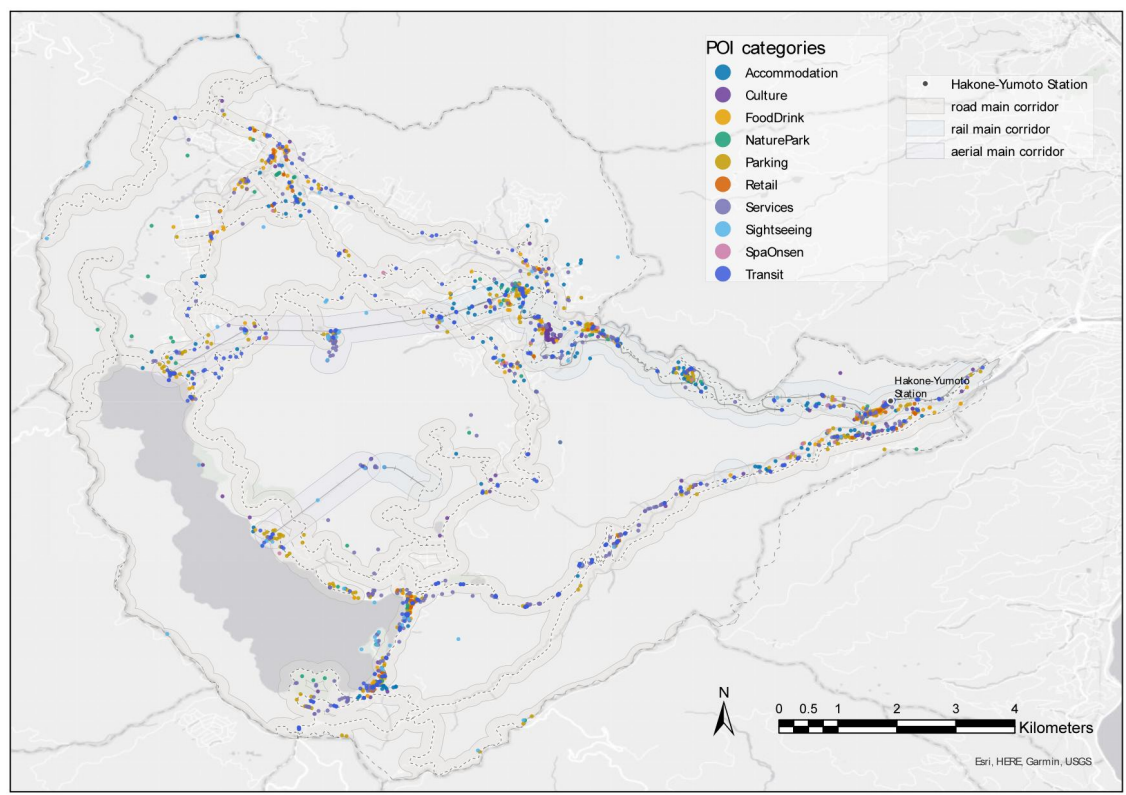

Fig. 1. Study area and POI categories in Hakone. The boundary defines the analysis domain for stay-event filtering and POI inventory curation. POIs are labeled by MID10 categories; main transport corridors and the Hakone–Yumoto Station are shown for orientation.

3.1.1 Stay-event data (human mobility)
We use high-density GPS trajectory data collected in November 2021 and convert raw trajectories into an ordered sequence of stay events for each individual-day. Each stay event record contains a user identifier, day label, start time and end time, start hour, dwell duration (minutes), and a representative event location (longitude/latitude). To support uncertainty-aware place semantics, each stay event is associated with a soft MID10 probability vector,

$$\mathbf{p}_t = (p_{t,1}, \ldots, p_{t,10}), \tag{1}$$

defined over the ten place-category semantics in MID10 (section 3.1.3). This soft semantic representation is used consistently throughout the pipeline, including for the train-side estimation of category transitions, dwell-time models, and weak spatial priors (sections 3.5–3.7).

3.1.2 POI inventory data (baseline and scenarios)
POIs are extracted from OpenStreetMap (OSM) and curated into a master POI table with stable identifiers. We retrieve and preprocess the OSM features using OSMnx, and then curate them into one baseline POI inventory and four scenario-specific inventories for counterfactual analysis. OSM data are distributed under the Open Database License; attribution to OpenStreetMap contributors is provided accordingly.

Each POI inventory table includes: (i) a stable POI ID ($poi\_id$), (ii) coordinates (longitude/latitude), and (iii) a MID10 category label. Scenario inventories are created by editing the baseline inventory while preserving stable POI identifiers wherever applicable. This allows for controlled counterfactual simulation in which behavioral parameters are held fixed and only the POI inventory is swapped (section 3.9).

3.1.3 MID10 semantics (place-category semantics)
We adopt a fixed MID10 taxonomy with the following categories and naming conventions, which are used consistently across all aggregation and visualization: Accommodation, Transit, Retail, Services, Parking, SpaOnsen, FoodDrink, Culture, Sightseeing, and NaturePark. A stay event's semantics are defined as place-category semantics (MID10) rather than latent "activity meaning" beyond the observed place category. The soft vector $\mathbf{p}_t$ represents uncertainty over MID10 semantics for each stay event and enables expected (probability-weighted) aggregation on real data.

3.2 Overview: event-chain simulator and its semi-Markov interpretation
We model tourist mobility as a time-ordered event chain where each element is a stay event characterized by

$$(h_t, c_t, d_t, \pi_t), \tag{2}$$

where $h_t$ is the stay hour, $c_t \in \{1, \ldots, 10\}$ is the MID10 category state, $d_t$ is the dwell time (minutes), and $\pi_t$ is an instantiated POI with attributes $(poi_id, lon, lat)$. The simulator is rhythm-consistent; that is, it is explicitly anchored to observed temporal signatures via (i) an hour–category start constraint, (ii) time-blocked category transitions, (iii) hour-dependent termination (stop hazard), and (iv) hour×category dwell distributions. Fig. 2 provides a full overview of the framework and the train-side artifacts used in the simulation. These rhythm signatures are planning-relevant because they summarize when different place categories are demanded (hour×category profiles) and how activity categories tend to evolve within a day (block-wise transitions), which directly supports time-windowed crowd management and category-specific service scheduling.

*Notation and discretization.* Each stay event $t$ is associated with a start time $s_t$ (timestamp) and a dwell duration $d_t$ (minutes). The start-hour variable $h_t \in \{0,1,\ldots,23\}$ is obtained by discretizing $s_t$ to its hour-of-day in local time. When a sampled dwell duration crosses an hour boundary, the "update hour" for the next-state transition is computed from the advanced clock time $s_t + d_t$ after applying the end-of-day truncation rule (section 3.4), and then discretized to $\{0,1,…,23\}$ in the same way.

Each stay event carries a soft MID10 semantic vector $\mathbf{p}_t = (p_{t,1}, …, p_{t,10})$, where $p_{t,c} \geq 0$ and $\sum_{c=1}^{10} p_{t,c} = 1$, representing uncertainty over the ten MID10 place-category semantics defined in section 3.1.3.

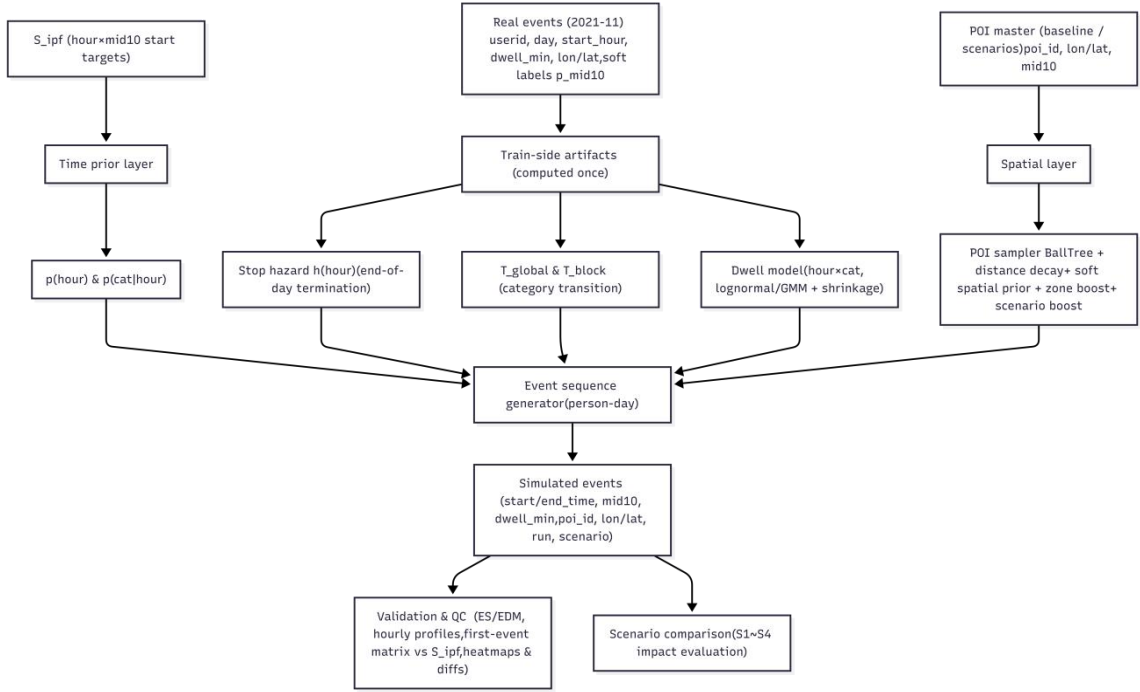

Fig. 2. Overview of the rhythm-consistent simulation framework. Train-side artifacts are computed once from observed stay events and reused across baseline and scenario runs, which swap only the POI inventory

*Model scope*. The simulator operates at the stay-event level and updates the mobility state only at event boundaries. Specifically, it generates a time-ordered sequence of stay events and associated POI instantiations but does *not* model inter-event travel time, produce continuous trajectories, and reconstruct point-level GPS paths. The model is designed for reproducing and comparing rhythm signatures (hour×category profiles, dwell distributions, and category transitions) under controlled POI-inventory changes.

*Semi-Markov requirements in our setting*. A (possibly time-inhomogeneous) semi-Markov process is characterized by (i) an embedded discrete state process, (ii) a sojourn-time distribution conditional on states (and potentially time), and (iii) a transition kernel applied at renewal (event) epochs. In our formulation, the embedded state is $c_t$, the sojourn time is $d_t$, and the transition kernel is time-block dependent $T^{(b)}$, where $b = b(h)$ maps an event time to a coarse time block (section 3.5). Termination is modeled separately via an empirical hour-dependent hazard $H(h)$ (section 3.4), which determines the end of the person-day chain.

*Operational definition of rhythm-consistency*. We consider the simulator rhythm-consistent if the Monte Carlo averages reproduce the observed hour×category Expected Stays (ES) and Expected Dwell Minutes (EDM) profiles under pre-specified similarity metrics (e.g., profile-wise root mean square error [RMSE] and Pearson correlation), while maintaining qualitatively consistent block-wise transition structure as measured by matrix-distance diagnostics on $T^{(b)}$ (section 3.10).

3.3 Start constraint from an hour×category target matrix $s_{ipf}$

Let $S_{ipf} \in \mathbb{R}^{24 \times 10}$ denote an hour×category target matrix for first stay events. $S_{ipf}$ is constructed as an hour×category target for first stay events by applying iterative proportional fitting to match the desired hour marginal and category marginal totals (Appendix A), yielding a matrix whose entries represent target counts (or masses) before conversion to the priors $p(h)$ and $p(c|h)$. We convert $S_{ipf}$ into a start-hour prior $p(h)$ and a conditional category prior $p(c|h)$:

$$p(h) = \frac{\sum_c S_{ipf}(h,c)}{\sum_{h'} \sum_c S_{ipf}(h',c)}, \quad p(c|h) = \frac{S_{ipf}(h,c)}{\sum_{c'} S_{ipf}(h,c')}. \qquad (3)-(4)$$

These priors initialize the first event as $h_0 \sim p(h)$ and $c_0 \sim p(c|h_0)$. Because $S_{ipf}$ is explicitly used as an initialization constraint, agreement with $S_{ipf}$ is treated as a constraint-compliance check rather than independent validation (section 3.10).

### 3.4 Hour-dependent termination: stop hazard $H(h)$

To reproduce realistic day lengths without imposing a fixed number of events, we estimate an empirical stop hazard $H(h)$ from real person-days. Let $h_{u,d}^{\max}$ be the start hour of the last stay event for user $u$ on day $d$. The discrete-time hazard is computed as

$$H(h) = Pr(h_{u,d}^{\max} = h | h_{u,d}^{\max} \geq h) = \frac{\#\{(u,d): h_{u,d}^{\max} = h\}}{\#\{(u,d): h_{u,d}^{\max} \geq h\}}. \qquad (5)$$

During simulation, after generating an event starting at hour $h_t$, the chain terminates with probability $H(h_t)$ (optionally scaled by a sensitivity multiplier; Appendix A). An explicit end-of-day cap is also applied to prevent events from crossing midnight.

### 3.5 Category transitions with time blocks and Dirichlet smoothing

We estimate the category-to-category transition tendencies using the soft semantic vectors $p_t$. For each consecutive event pair $(t, t+1)$, we add an expected outer-product count $p_t p_{t+1}^\top$. Aggregating over all sequences yields an expected transition count matrix $N$. Dirichlet smoothing with parameter $\alpha$ is applied elementwise before row-normalization:

$$T(i,j) = \frac{N(i,j) + \alpha}{\sum_{j'}(N(i,j') + \alpha)}. \qquad (6)$$

To capture within-day rhythm changes, we use a piecewise-constant time-inhomogeneous approximation by defining coarse time blocks $b$ (Appendix A) and estimating block-specific kernels $T^{(b)}$ by restricting the outer-product counts to transitions the current event hour of which falls in block $b$. In the simulation, the next category is sampled as $c_{t+1} \sim T^{(b(h_{t+1}))}(c_t, \cdot)$, where $h_{t+1}$ is the update hour after advancing time by the sampled dwell duration.

### 3.6 Dwell-time model $p(d|h,c)$: mixture on log scale with shrinkage

We model the stay duration $d_t$ (minutes) conditional on start hour $h_t$ and category $c_t$. Empirically, dwell times are heavy-tailed and often exhibit multi-modality (e.g., short vs. long stays). Therefore, we fit a two-component Gaussian mixture to $\log d$ at the hour×category level when the effective weighted sample size is sufficient; otherwise, we revert to a category-level global model.

For each $(h,c)$, we use weights $w_n = p_{n,c}$ from the soft labels of real stay events starting at hour $h$ and define the effective weight $W_{eff} = \sum_n w_n$. If $W_{eff}$ is below a threshold (Appendix A), we use the category-level global model for that hour. When a local mixture is fit, we apply a simple convex shrinkage toward the corresponding global parameters to improve stability under sparse $(h,c)$ cells. We enforce a minimum dwell duration and truncate events at the end-of-day cap (Appendix A). Implementation-level thresholds and fallback rules are summarized in Appendix A for reproducibility.

### 3.7 POI inventory and likelihood-based probabilistic event-to-POI assignment
#### 3.7.1 POI master tables and scenario interventions

We maintain (i) one baseline POI inventory and (ii) four scenario inventories, each stored as a POI master table with stable IDs and MID10 labels. Scenario construction follows four intervention types (type switch, new POI addition, relocation via coordinate swap, and removal), summarized in Table 3.1.

Across all runs, the behavioral artifacts (start priors, transition kernels, dwell-time models, and stop hazard) are held fixed, and scenario effects are driven primarily by swapping the POI inventory. In the main experiments, we additionally apply a controlled changed-POI emphasis that upweights POIs the IDs/categories of which differ from the baseline in order to improve intervention detectability under finite Monte Carlo sampling (Appendix A); we report robustness checks with this emphasis disabled. This design enables controlled counterfactual comparison under POI supply changes. In the released simulator implementation used for the main experiments, POI selection additionally applies (i) a scenario-aware emphasis for POIs altered by the counterfactual inventory and (ii) a soft Hakone–Yumoto zone rebalancing derived from observed stay-event matches; both are treated as bias controls for spatial anchoring and scenario observability rather than causal behavioral effects (Appendix A).

Table 1. POI configuration change types used for counterfactual scenario design and expected impacts.

| POI change type | What changes | Counterfactual setup | Expected sequence shift | Common metrics |
|---|---|---|---|---|
| Type switch (same location) | Category A → B | Change label/weights only | Category share; local transitions | ΔES, ΔEDM, ΔT_h |
| New POI (new location) | Add POI node | Add one POI, keep others fixed | New hotspot; redistribution | ΔES, ΔEDM, ΔT_h, ΔHit |
| Relocation | Move POI coordinates | Pairwise coordinate swap (category fixed) | Hotspot shift; access-driven transitions | ΔES, ΔEDM, ΔT_h, ΔHit |
| Removal | Delete POI | Remove node / set weight = 0 | Substitution to nearby same-class | ΔES, ΔEDM, ΔT_h |

Notes: ES = Expected Stays (hour×category), EDM = Expected Dwell Minutes (hour×category), ΔT_h = change in hour-specific transition tendency, ΔHit = map hit-rate within a reference radius (e.g., 100–200 m) when a spatial anchor is required.

3.7.2 Candidate retrieval via BallTree (haversine)

Given an event location $(lon_e, lat_e)$ and a target category $c$, we retrieve a candidate set $Z_c(e)$ using a BallTree built on POI coordinates with the haversine metric. By default, we query the $K$ nearest POIs within the same category. With a small probability, an exploration move samples candidates within a fixed radius to reduce local trapping under highly clustered POIs (Appendix A). If category-specific retrieval is unavailable (e.g., empty category inventory), we fall back to a global nearest-neighbor query.

3.7.3 Prior-likelihood weighting and posterior-like normalization

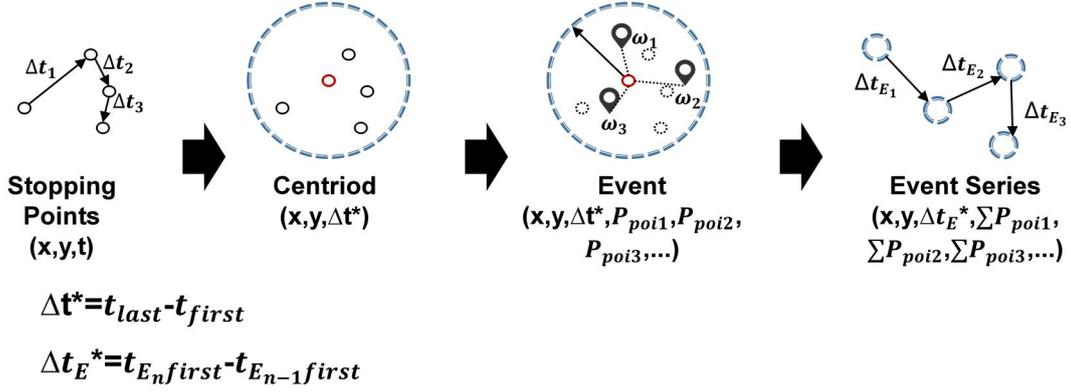

$\Delta t^* = t_{last} - t_{first}$

$\Delta t_E^* = t_{E_n first} - t_{E_{n-1} first}$

Fig. 3. Conceptual illustration of candidate POI retrieval and posterior-like weighting. A stay-event centroid defines a candidate set within a category; relative weights combine distance-decay likelihood and weak prior information before normalization and sampling.

In the event-chain simulator (Fig. 3), each simulated stay event is instantiated to a concrete POI by assigning a probability distribution over a candidate set $Z_c(e)$, retrieved for the event location and target category (section 3.7.2). We compute a posterior-like score for each candidate POI $i \in Z_c(e)$ through an explicit prior-likelihood weighting,

$$s_i(e) \propto L(e|i) \cdot \pi(i) \cdot m_i, \quad (7)$$

and normalize it within the candidate set to obtain a well-defined sampling distribution:

$$P(i|e) = \frac{s_i(e) + \varepsilon}{\sum_{k \in Z_e(e)} (s_k(e) + \varepsilon)}. \quad (8)$$

We set $\varepsilon = 10^{-12}$, which is orders of magnitude smaller than typical nonzero scores and, therefore, does not materially change the relative weights; it only prevents undefined 0/0 cases when all candidate scores collapse to zero. Here, $\varepsilon > 0$ is a small stabilizer to prevent 0/0 and numerical collapse under extreme sparsity; scores are normalized within the candidate set $Z_c(e)$ using this $\varepsilon$-stabilized denominator (Appendix A). The likelihood term is approximated by a distance-decay kernel:

$$L(e|i) = \exp\left(-\left(\frac{d(e,i)}{R_c}\right)^\gamma\right), \quad (9)$$

where $d(e,i)$ is the great-circle distance and $R_c$ is a category-specific distance scale (Appendix A). Here, the likelihood term should be understood as a relative geographic matching-strength proxy derived from great-circle distance than as a calibrated GPS measurement-error model. Accordingly, the prior-likelihood score is a heuristic Bayesian-style weighting that yields a normalized, uncertainty-aware sampling distribution over candidates.

In this implementation, the prior $\pi(i)$ represents a weak, data-derived preference over the POIs accumulated from soft-matched observed stay events; it is used to stabilize sampling and is not interpreted as a causal utility. The prior $\pi(i)$ is constructed as weak information by accumulating soft evidence from real stay events: each real stay event is matched to its nearest POI within a radius, and its soft MID10 weights are added to POI-level prior counters. The resulting prior is used only to stabilize sampling and reflect recurrent attractiveness patterns in observed data; it is not intended for causal interpretation (e.g., it should not be read as an intrinsic "utility" of the POI). The optional factor $m_i$ is used only for scenario emphasis/sensitivity experiments (Appendix A).

*Prior handling under scenario edits*. The weak prior $\pi(i)$ is computed once from the observed stay events on the baseline inventory and is used only as a stabilizer in sampling. Under scenario modifications, prior values are handled by deterministic bookkeeping rules to avoid introducing discretionary bias: (i) added POIs are initialized with a small baseline mass (category-level minimum prior) before normalization within the candidate set; (ii) removed POIs are assigned zero mass and are excluded from candidate retrieval, after which candidate-set normalization is applied; and (iii) for type-switch POIs that preserve the same POI identifier and location, the prior mass is retained at the POI level but is only active when the POI belongs to the retrieved category-specific candidate set. These rules ensure that scenario effects are driven by inventory changes and the likelihood term, whereas the prior contributes only weak regularization.

Consistent with this design, we describe the assignment conservatively as follows: We perform a likelihood-based probabilistic event-to-POI assignment using an explicit prior-likelihood weighting, which results in a posterior-like score over candidate POIs. This is a pragmatic formulation for uncertainty-aware matching and is not intended to be a fully specified generative model of GPS errors.

3.8 Event-chain generation algorithm (person-day)

Given a simulated person-day, we generate an ordered sequence of stay events. We first sample the initial hour and category from the $s_{ipf}$-derived priors (section 3.3). At each step $t$, we (i) sample dwell time $d_t \sim p(d|h_t, c_t)$ (section 3.6), (ii) assign a POI $\pi_t \sim P(i|e)$ using the prior-likelihood weighting (section 3.7), and (iii) advance time deterministically by adding $d_t$ to the current clock (truncated at end-of-day if needed). We then apply the stop hazard $H(h_t)$ as an hour-dependent termination probability (section 3.4). If the chain continues, we sample the next category from the appropriate time-block transition kernel $T^{(b)}$ (section 3.5) and repeat. Algorithm 3.1 summarizes the person-day generation procedure.

*Algorithm 3.1 (Person-day event-chain generation).* Initialize $(h_0, c_0) \sim p(h)p(c|h)$. For $t = 0,1…$until termination: sample $d_t \sim p(d|h_t, c_t)$; sample $\pi_t \sim P(i|e)$ from the candidate set; record $(h_t, c_t, d_t, \pi_t)$ and advance time; terminate with probability $H(h_t)$ or when reaching the end-of-day cap; otherwise sample $c_{t+1} \sim T^{(b(h_{t+1}))}(c_t, \cdot)$ and continue.

3.9 Monte Carlo simulation and scenario execution

We separate computation into (i) train-side artifacts computed once from real stay events and (ii) scenario runs that only swap the POI inventory. Train-side artifacts include the stop hazard $H(h)$, global and block-specific transition kernels $T$ and $T^{(b)}$, dwell-time models $p(d|h,c)$, and $S_{ipf}$-derived start priors. For scenario analysis, we sequentially run the simulator under each scenario POI inventory but keep all behavioral artifacts fixed. To enable fair comparisons across scenarios, we use consistent Monte Carlo settings (number of simulated users, runs, and seed control; Appendix A) and output both the event-level simulation logs and aggregated matrices used for evaluation. Across scenarios, paired comparisons are enforced by resetting the random number generator(RNG) seed per scenario so that stochastic draws are aligned; thus, any differences are attributable to the POI inventory swap and the associated POI selection bias controls but not Monte Carlo randomness (Appendix A).

3.10 Validation and quality control

To avoid circular validation, we report two classes of checks.

*Constraint compliance checks (not independent validation).* Because $S_{ipf}$ is explicitly used to initialize the first event, we only evaluate first-event start matrices as a compliance check. Specifically, we compute the simulated first-event hour×category matrix and report its relative Frobenius deviation from $S_{ipf}$. This verifies that the initialization constraint is respected but is not treated as independent evidence of model validity.

*Out-of-constraint validation.* We validate rhythm reproduction using aggregates that are not directly enforced as hard constraints at every step. We compute (i) ES as event counts aggregated by hour×category and (ii) EDM as total dwell minutes aggregated by hour×category, and compare the observed vs. simulated profiles. Additional diagnostics include the similarity of the transition structure between the observed and simulated category transitions and the spatial plausibility summaries derived from the simulated POI selections (e.g., shares within a reference zone when relevant). These evaluations quantify whether the simulator preserves key rhythm signatures beyond simple start-constraint compliance. These rhythm signatures are planning-relevant because hour×category ES/EDM profiles characterize time-windowed demand by place function (supporting staffing, crowd management, and service scheduling), whereas transition structure indicates how category-level activity sequences propagate within the day under supply perturbations.

## 4. Results
4.1 Rhythm consistency validation (baseline realism)

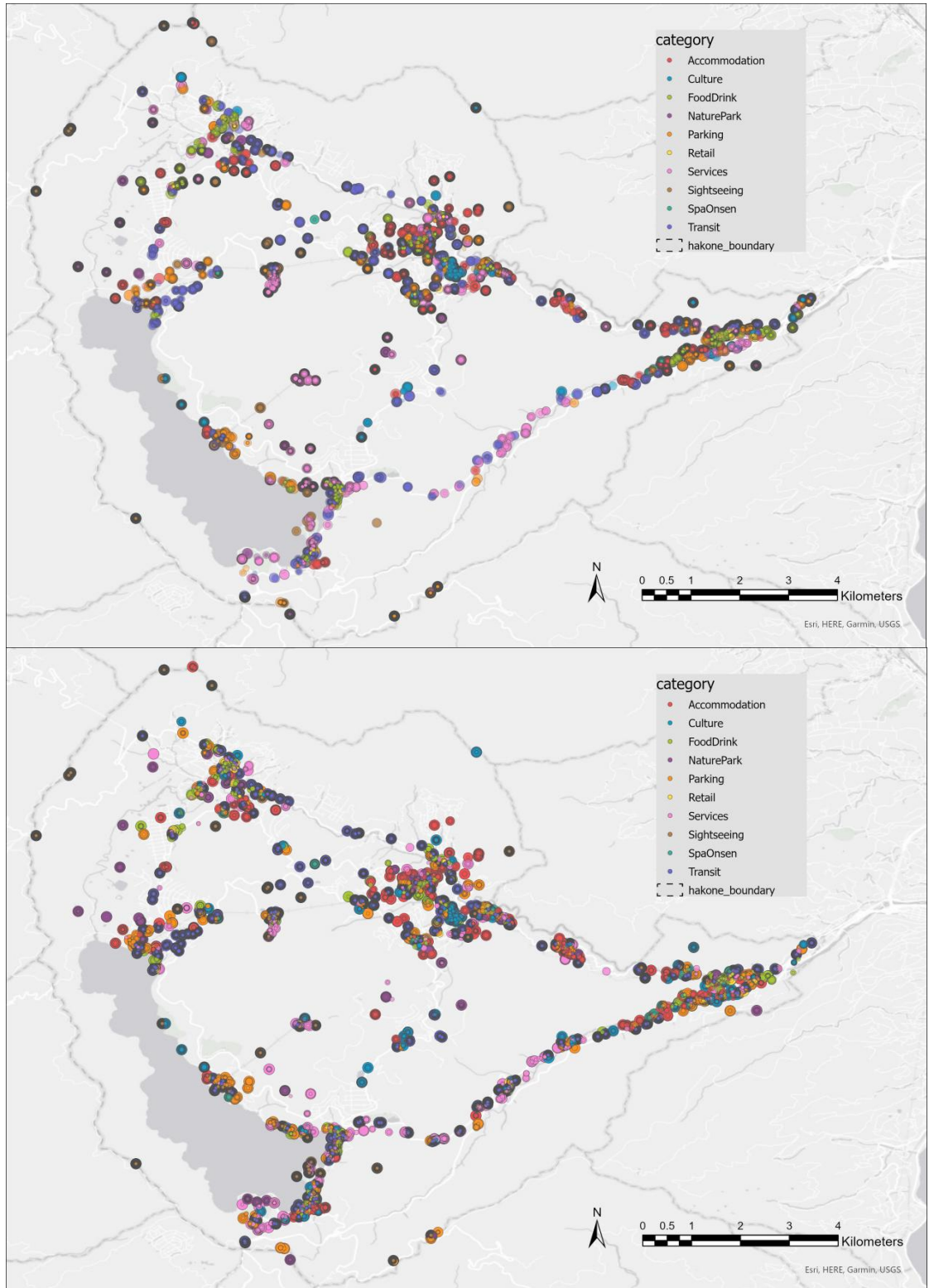

Fig. 4. Spatial footprint of observed vs. simulated stay events (MID10-colored points) in Hakone (Nov 2021 baseline).

We first assess whether the baseline simulator reproduces the spatial footprint and diurnal rhythm of the stay activity under the fixed analysis boundary and MID10 semantics.

*Spatial footprint*. Fig. 4 contrasts observed stay-event locations (Nov 2021) and simulated stay-event locations under the baseline configuration. The simulated point cloud reproduces the major spatial corridors and high-density clusters, while maintaining a comparable MID10 category composition at prominent hotspots. This provides an intuitive check that the probabilistic event-to-POI instantiation does not collapse activity into a small subset of locations and that the spatial anchors remain consistent with the empirical domain.

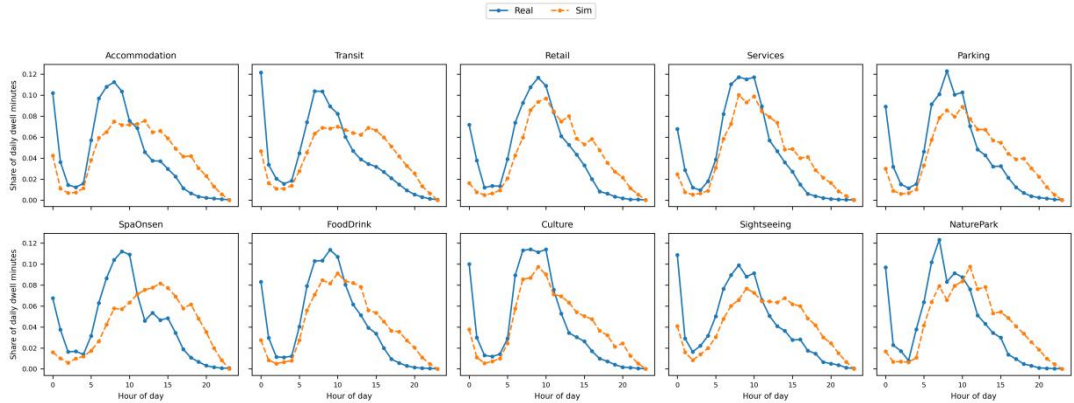

Fig. 5. Category-wise diurnal profiles of normalized dwell minutes (EDM share) for observed vs. simulated events.

*Diurnal rhythm at the category level (EDM)*. Fig. 5 compares observed vs. simulated category-wise diurnal profiles of normalized dwell minutes (EDM share). Across MID10 categories, the simulator reproduces the characteristic time-of-day structure (e.g., the morning-to-midday concentration in FoodDrink/Retail/Services and category-specific afternoon decay). This supports the intended rhythm-consistent property: the simulation preserves aggregate temporal regularity while generating discrete event chains with explicit dwell durations. To quantify profile agreement beyond visual comparison, Table 4.1 reports category-wise RMSE and Pearson correlation between observed and simulated EDM-share diurnal profiles, together with the macro- and EDM-weighted summaries.

Table 2. Similarity metrics for MID10-specific diurnal profiles of EDM share (observed vs. simulated).

| MID 10 | EDM RMSE | EDM Pearson r | EDM weight |
|---|---|---|---|
| Accommodation | 0.0311 | 0.583 | 933,465 |
| Transit | 0.0307 | 0.507 | 7,218,251 |
| Retail | 0.0273 | 0.702 | 323,382 |
| Services | 0.0241 | 0.818 | 1,048,368 |
| Parking | 0.0270 | 0.718 | 1,093,066 |
| SpaOnsen | 0.0373 | 0.279 | 101,318 |
| FoodDrink | 0.0259 | 0.744 | 807,036 |
| Culture | 0.0271 | 0.765 | 1,290,000 |
| Sightseeing | 0.0286 | 0.533 | 1,994,520 |
| NaturePark | 0.0310 | 0.600 | 414,506 |

For each MID10 category, we compute a 24-hour vector of normalized EDM share (hourly dwell minutes divided by the category's total dwell minutes over the evaluation period) for both the observed and simulated data. The evaluation period is November 2021 under the fixed analysis boundary. We report the RMSE and Pearson correlation (r) between the two

24-hour share vectors. EDM weight denotes the observed category total dwell minutes used for weighted averaging. Macro-averages treat categories equally; EDM-weighted averages weight by observed category total dwell minutes. Overall, the macro-averaged Pearson correlation (r) of category-wise EDM-share diurnal profiles is 0.625, while the EDM-weighted Pearson correlation (r) is 0.591; the corresponding macro-averaged RMSE is 0.0290 and the EDM-weighted RMSE is 0.0292. The largest discrepancy occurs for SpaOnsen (Pearson r = 0.279, RMSE = 0.0373), which is expected given its low EDM mass (EDM weight = $1.01 \times 10^5$, an order of magnitude smaller than major categories such as Transit or Sightseeing) and its high temporal heterogeneity, driven by facility operating patterns and longer, more variable dwell durations; under such sparsity, small absolute timing shifts can substantially reduce correlation even when the overall contribution to total EDM is minor. This is also consistent with the gap between macro-averaged and EDM-weighted summaries, where low-mass categories contribute less to the weighted agreement.

4.2 Hour-by-day rhythm diagnostics (fine-grained heatmaps)

After the coarse diurnal profiles, we evaluate whether the simulator preserves fine-grained rhythm structure across the month using day×hour heatmaps. We then diagnose the systematic residual patterns.

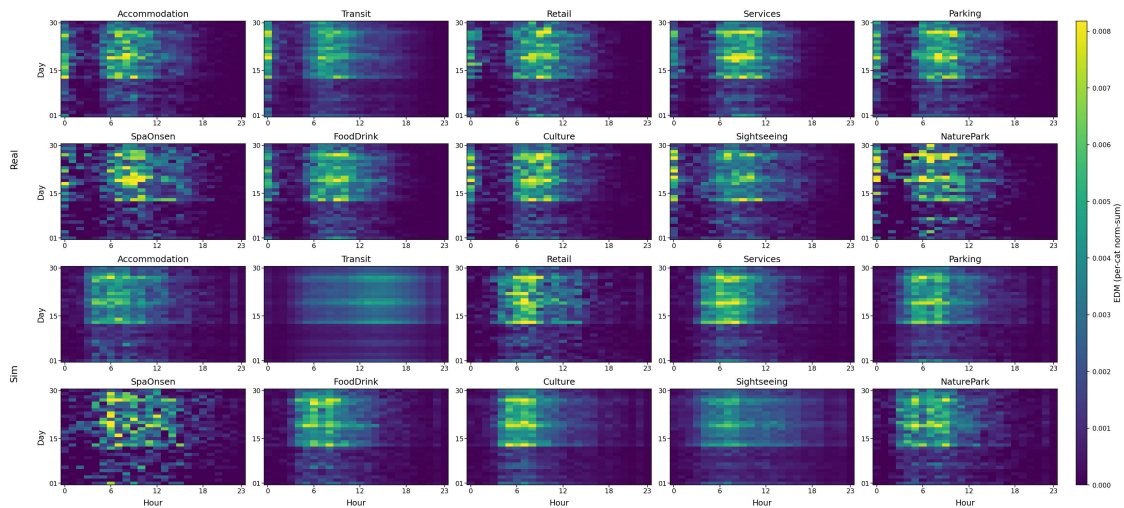

Fig. 6. Day-by-hour heatmaps of normalized EDM (per-category normalized sum): observed vs. simulated.

*Observed vs. simulated heatmaps (EDM)*. Fig. 6 reports the day-by-hour heatmaps of the normalized EDM (per-category normalized sum) for the observed and simulated data. The simulation reproduces the dominant within-day concentration window and its persistence across days for most categories, indicating that the model's hour-conditioned dwell mechanism and time-block transitions jointly preserve the month-level rhythm signature beyond a single daily average curve.

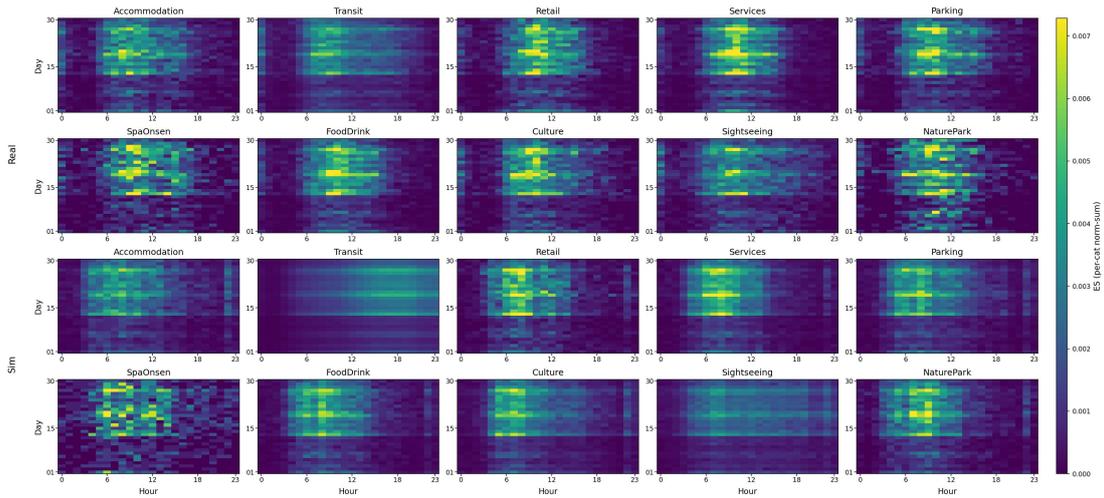

Fig. 7. Day-by-hour heatmaps of normalized ES (per-category normalized sum): observed vs. simulated.

*Observed vs. simulated heatmaps (ES)*. The ES heatmaps are per-category normalized to sum to 1 across all day×hour cells: for each MID10 category, the day×hour counts are divided by the category's total ES over the evaluation period. Therefore, residuals represent absolute deviations in share mass, and not raw event counts. Fig. 7 repeats the comparison for ES (event counts). Consistency between the EDM and ES heatmaps is important because EDM integrates both event frequency and dwell durations; agreement in both metrics suggests the simulator is not matching dwell minutes by compensating errors in event counts (or vice versa) but preserving the rhythm structure in a coherent manner.

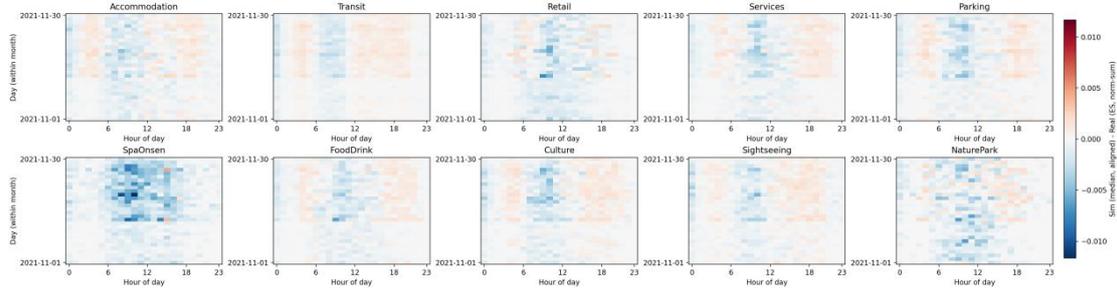

Fig. 8. Residual heatmaps for ES (phase-aligned simulation median minus observation), highlighting systematic deviations.

*Residual structure (phase-aligned)*. To identify systematic deviations, Fig. 8 visualizes phase-aligned residual heatmaps (simulation median over runs minus observation) for ES. Residuals concentrate in specific hour windows and categories, revealing where the simulator under- or over-represents event intensity even if the overall rhythm is broadly reproduced. These residual patterns motivate the scenario section: scenario impacts should be interpreted relative to the (i) baseline residual magnitude and (ii) spatial anchoring/bias controls described in Appendix A. To summarize residual amplitudes on the normalized-share scale, Table 4.2 reports the mean absolute residual (MAR), 95th percentile of absolute residual (P95), and Frobenius norms of ES residual heatmaps by category, together with macro- and ES-weighted summaries.

Table 3. Residual-amplitude metrics for ES day×hour heatmaps (simulation minus observation).

| MID10 | ES resid MAR | ES resid P95 | ES resid Fro | ES weight |
|---|---|---|---|---|
| Accommodation | 0.000897 | 0.002456 | 0.0321 | 2,708 |
| Transit | 0.000792 | 0.002127 | 0.0284 | 24,606 |
| Retail | 0.000744 | 0.002034 | 0.0279 | 1,202 |
| Services | 0.000727 | 0.002136 | 0.0276 | 3,585 |
| Parking | 0.000818 | 0.002316 | 0.0304 | 3,623 |
| SpaOnsen | 0.001147 | 0.003447 | 0.0444 | 336 |
| FoodDrink | 0.000757 | 0.002045 | 0.0282 | 2,831 |
| Culture | 0.000858 | 0.002511 | 0.0318 | 3,825 |
| Sightseeing | 0.000846 | 0.002308 | 0.0306 | 6,526 |
| NaturePark | 0.000964 | 0.002806 | 0.0363 | 1,691 |

ES heatmaps are per-category normalized to sum to 1 across all day×hour cells for each MID10 category. For each category, we compute residuals on this normalized-share scale and report: MAR, P95, and Frobenius norm of the residual matrix. The ES weight is the observed category total ES used for weighted averaging. Overall, the residual amplitudes are small on the normalized-share scale: the macro-averaged MAR is $8.55 \times 10^{-4}$ (≈0.086% absolute share mass per cell on average), and the 95th-percentile residual is $2.42 \times 10^{-3}$ (≈0.24%). The corresponding ES-weighted MAR is $8.12 \times 10^{-4}$, indicating that deviations remain small even when emphasizing high-volume categories. SpaOnsen shows the largest residuals among the categories, consistent with a weaker rhythm reproducibility for that category. Beyond heatmap residuals, we also test whether the simulator preserves the underlying time-block transition structure that governs how categories connect across events.

Table 4. Distances between the observed time-block transition matrices and transition matrices re-estimated from simulated sequences. n_transitions denotes the number of adjacent event pairs used for estimating the transition matrix in each time block.

| block | Block start | Block end | N transitions | Frobenius dist | Cosine sim | Rowwise JS mean |
|---|---|---|---|---|---|---|
| 0 | 0 | 4 | 27,919 | 0.177 | 0.993799 | 0.003022 |
| 1 | 5 | 7 | 57,021 | 0.190 | 0.993743 | 0.002656 |
| 2 | 8 | 10 | 99,842 | 0.235 | 0.991116 | 0.003739 |
| 3 | 11 | 13 | 106,301 | 0.264 | 0.988657 | 0.004681 |
| 4 | 14 | 17 | 136,378 | 0.308 | 0.984264 | 0.005457 |
| 5 | 18 | 23 | 82,863 | 0.0654 | 0.999355 | 0.000633 |

Because the simulator's rhythm is driven not only by start-time priors and dwell models but also by time-block transitions, we examine whether the simulated event sequences reproduce the empirical transition structure at the same time-block resolution. To avoid circular validation, we do not compare the observed transition matrices to the transition kernels injected into the simulator. Instead, we re-estimate the time-block transition matrices from the simulated sequences using the same estimator as in the observed data, and then compare these re-estimated matrices to the observed ones (Table 4.3).

Across blocks, the re-estimated matrices remain structurally close to the observed matrices (cosine similarity 0.984-0.999 and small row-wise Jensen-Shannon divergence means on the order of $10^{-3}$), while nonzero Frobenius distances indicate that the simulator is not trivially reproducing the input kernels. Deviations are most pronounced in the afternoon block (14-17; Frobenius = 0.308, cosine = 0.984), whereas the evening block (18-23) shows the tightest agreement (Frobenius = 0.065, cosine = 0.999). This pattern suggests that midday transitions are harder to recover from generated sequences, likely because activity types are more mixed and destination choices compete more strongly, whereas late-day transitions are more stereotyped and, therefore, more stable under simulation.

4.3 Scenario impacts (spatial redistribution of ES/EDM)
We now quantify how counterfactual POI interventions reshape the spatial distribution of stay intensity. Throughout, behavioral artifacts (start priors, transition kernels, dwell models, stop hazard) are held fixed; scenario effects are driven by swapping the POI inventory and the associated POI-selection controls (Appendix A). Importantly, these counterfactual results should be interpreted as model-based redistribution under fixed behavioral kernels, not as causal estimates of real-world policy effects. The analysis is designed for relative comparisons between scenarios under controlled assumptions, which helps identify planning-sensitive locations and categories while keeping the behavioral mechanism transparent. We report impacts for both EDM (minutes) and ES (events) because they reflect different planning-relevant quantities (time spent vs. visit frequency).

To make scenario impacts interpretable for street-level planning, we focus on FoodDrink and Retail as target MID10 types because they (i) have direct planning relevance in typical street revitalization (e.g., introducing convenience stores/coffee shops/specialty stores, or closing/relocating a store), and (ii) jointly influence both dwell minutes and the common activity chain structure (e.g., transitions resembling Transit → FoodDrink → Sightseeing in daily sequences).

Target POIs are chosen within MID10 FoodDrink/Retail, emphasizing locations with frequent matched events and large cumulative dwell minutes (dwell_sum). We restrict candidates to the Hakone–Yumoto shopping street main corridor (or an important branch) to maintain planning plausibility, and we further require a mixed-POI neighborhood within ~200 m to enable realistic substitution/redistribution effects in scenario simulations.

*Selected POIs (roles used in scenario bookkeeping)*. The scenario design uses a small set of explicitly identified POIs with role tags (e.g., target_food, target_retail_for_swap, and a newly introduced POI such as S2_new_fooddrink), each defined by a stable poi_id, MID10 label, and coordinates (lon/lat). These identifiers ensure that "what changed" is traceable and that scenario-minus-baseline maps can be attributed to specific POI edits rather than diffuse inventory drift.

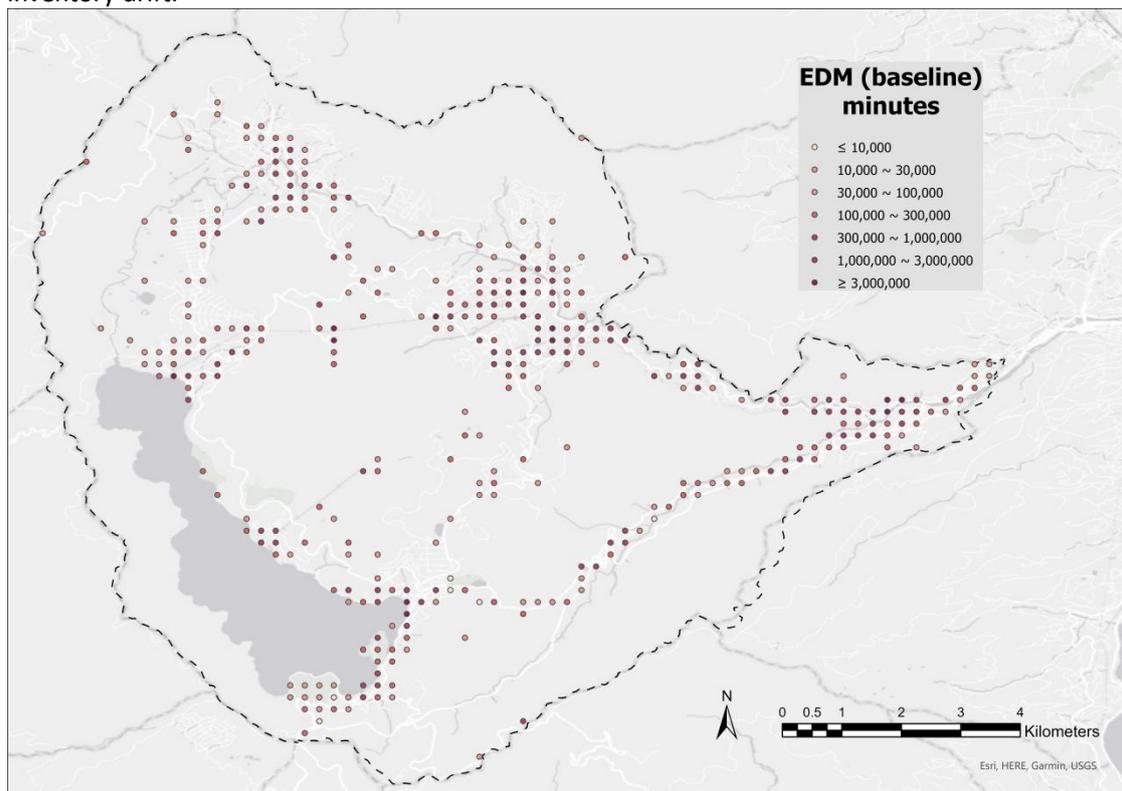

Fig. 9. Baseline spatial distribution of aggregated dwell minutes (EDM) over the analysis grid.

Fig. 9 maps the baseline spatial distribution of aggregated dwell minutes (EDM). High-intensity grids align with the major activity corridor and prominent clusters, providing the reference against which scenario-induced redistribution is evaluated. Consistent baseline anchoring is critical because the scenario maps are interpreted as redistribution around the same corridor structure, and not as a redefinition of the study boundary or aggregation grid.

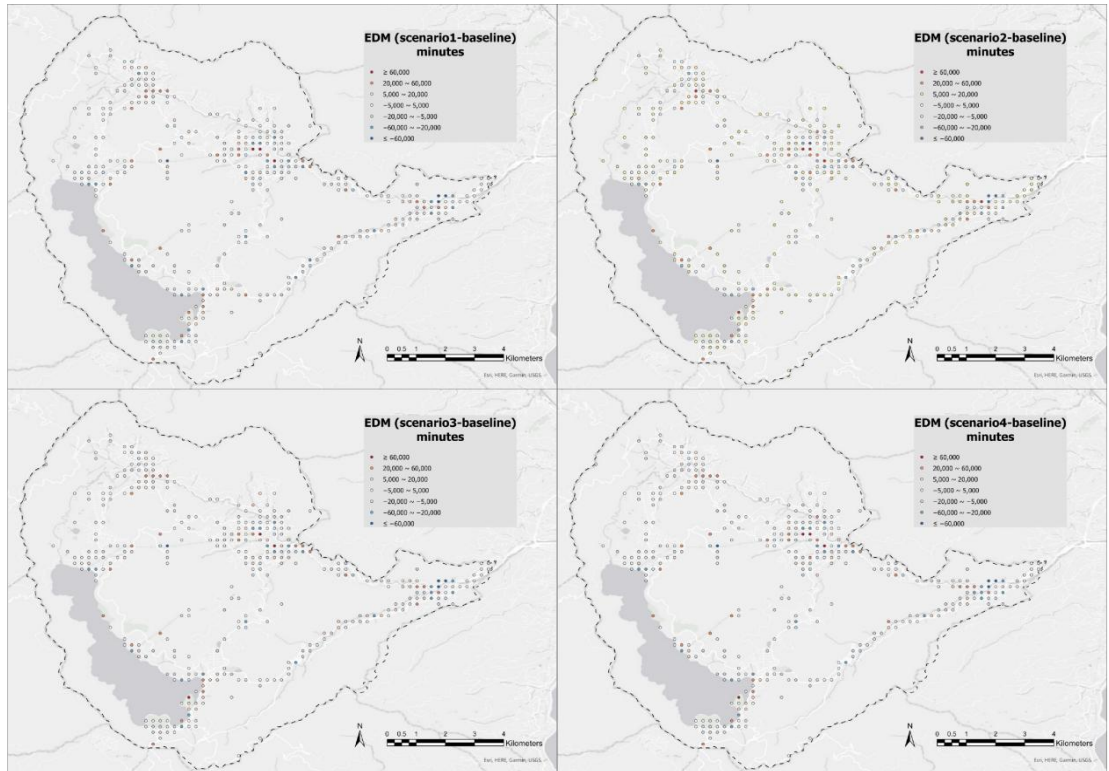

Fig. 10. Scenario-induced spatial changes in EDM (scenario minus baseline) for S1–S4.

Fig. 10 visualizes the EDM differences (scenario minus baseline) for S1–S4. The difference maps highlight where interventions increase or decrease dwell minutes locally and along the corridor. The EDM changes reflect both (i) where events are instantiated and (ii) hour-conditioned dwell-time consequences of category composition; therefore, the EDM maps directly indicate time-based planning impacts (e.g., crowding pressure measured in person-minutes).

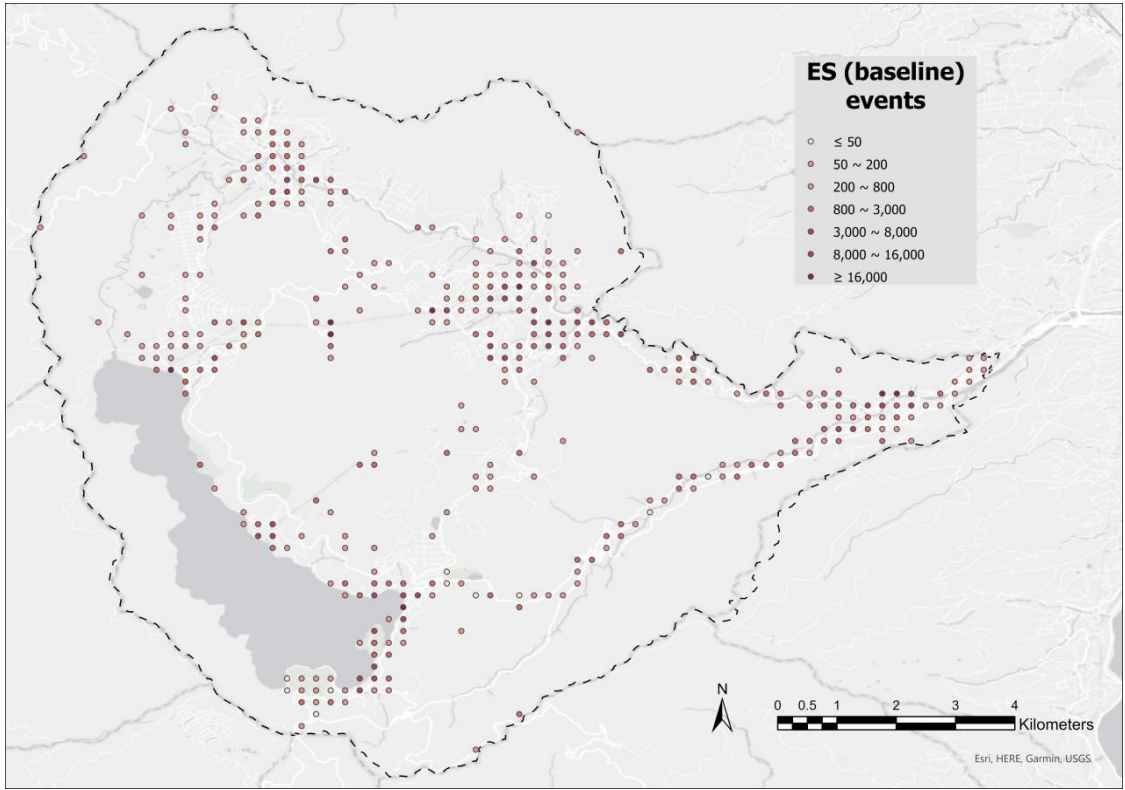

Fig. 11. Baseline spatial distribution of event counts (ES) over the analysis grid.

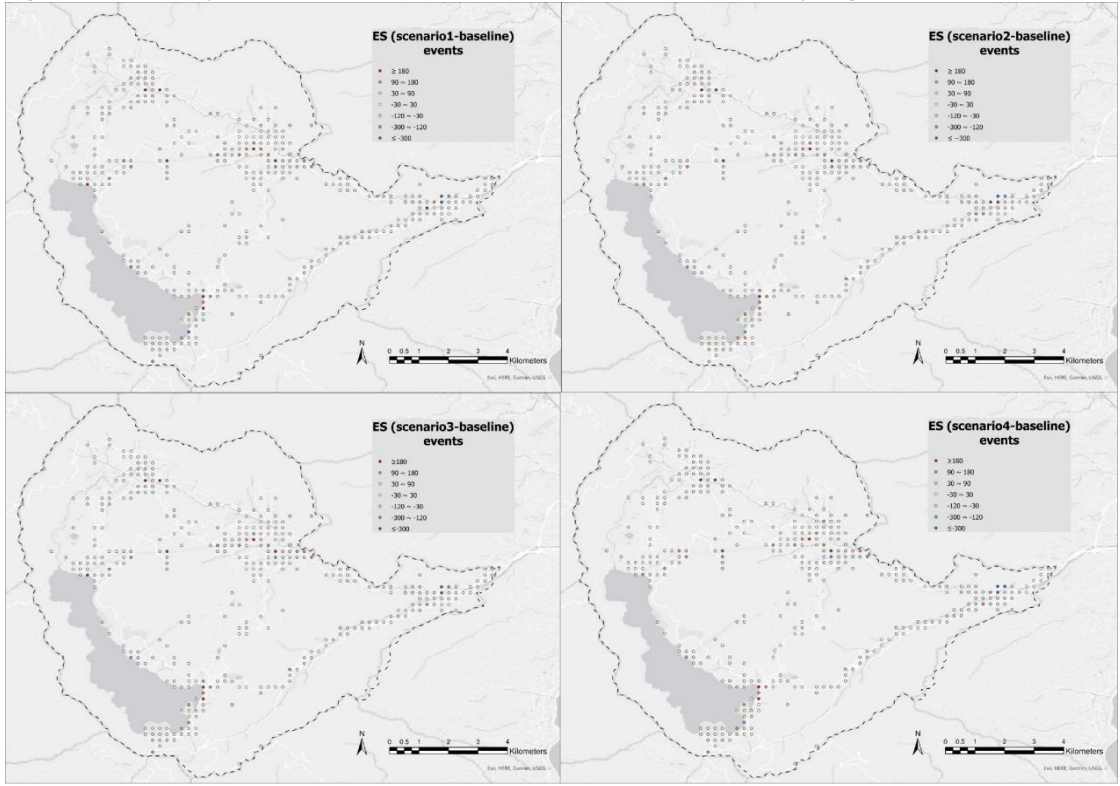

Fig. 12. Scenario-induced spatial changes in ES (scenario minus baseline) for S1–S4.

Fig. 11 maps the baseline ES (event counts) as a complementary reference. Fig. 12 then reports the ES differences (scenario minus baseline) for S1–S4. Comparing Fig. 10 and 12 clarifies whether a scenario primarily shifts visit frequency (ES) or time spent (EDM). A scenario may increase the ES with limited EDM change when it induces more short stops, whereas EDM-dominant shifts indicate longer stays and/or substitutions into longer-dwell categories.

## 5. Conclusion

This study addresses the gap between tourism and urban planning. Although GPS trajectories are becoming increasingly available, they are noisy, irregularly sampled, and weakly linked to place semantics, whereas planning questions are scenario-oriented and require interpretable, spatially explicit evidence. We proposed an event-based simulation framework that represents tourist mobility as sequences of stay events with probabilistic event-to-POI anchoring and evaluated POI inventory interventions through scenario runs that keep behavioral components fixed.

The results support three main conclusions. First, the framework yields a planning-relevant representation that is consistent with the observed behavior at both the spatial and temporal scales. Spatially, the simulated stays reproduced the observed concentration of activity along the main destination corridor and around key hubs, without introducing implausible new hotspots. Temporally, the simulator recovered the diurnal allocation of dwell minutes across the MID10 categories, indicating that category-specific diurnal allocations of dwell minutes were preserved rather than matching totals alone. Taken together, these checks indicate that the simulator retains the principal anchors of the destination that matter to management and design decisions.

Second, the simulator reproduced a fine-grained temporal structure beyond the aggregate profiles. When comparing day-by-hour patterns, the simulation aligned with the observed timing and intensity across categories. Residuals are concentrated in specific-hour windows rather than being uniformly distributed across the day×hour grid, suggesting that the simulator captures fine-grained temporal regularities relevant to time-targeted operations. This is practical because it suggests that the model captures the temporal regularity required for operational questions such as time-targeted congestion mitigation, staffing, and phased interventions.

Third, scenario experiments show that the framework can serve as a coherent counterfactual engine for POI-oriented interventions. With a fixed behavioral kernel and only POI supply changes, the resulting spatial redistribution remains geographically plausible at the destination scale. Reporting both event counts (ES) and dwell minutes (EDM) reveals complementary mechanisms; some interventions primarily shift how many stays occur, whereas others shift how long the stays last. This distinction is directly relevant, because capacity pressure and congestion risks often track event intensity, whereas economic exposure and experience design are frequently more sensitive to dwell-time redistribution.

Overall, this study contributes to an end-to-end reproducible pipeline that bridges raw mobility traces and scenario-based planning requirements. Methodologically, it formalizes tourist mobility as an event chain with state-dependent dwell and time-conditioned transitions while maintaining an uncertainty-aware linkage to POIs. Empirically, this approach can reproduce the observed spatial footprints and temporal allocations without scenario-specific refitting, enabling transparent comparisons across POI interventions.

The current framework abstracts inter-event travel time and does not condition transitions or dwell on exogenous covariates (e.g., weather or capacity). Future work could introduce user heterogeneity and covariate-conditioned kernels while preserving the same validation logic and scenario comparability.

**CRediT authorship contribution statement**

**Jianhao Shi:** Formal analysis, Investigation, Validation, Visualization, Writing – original draft. **Tomio Miwa:** Conceptualization, Methodology, Funding acquisition, Supervision, Writing – review & editing. **Wanglin Yan:** Resources, Data curation, Funding acquisition.

**Declaration of competing interest**

The authors declare no competing interests.

**Data availability**

**Appendix A. Reproducibility: implementation settings and reporting checklist**

This appendix summarizes the implementation-level settings required to reproduce the simulation results and clarifies which quantities are (i) fixed hyperparameters, (ii) estimated from observed stay events, and (iii) optional sensitivity knobs. Unless stated otherwise, the main results use the core settings in Table A.1. To avoid ambiguity, Table A.1 also lists the POI-selection bias controls (scenario boost and Yumoto zone boost) that are enabled in the released simulator implementation and, therefore part, of the main experiment configuration.

*Estimated artifacts (not "defaults")*. Transition kernels $T^{(b)}$, dwell-time parameters, and the stop hazard $H(h)$ are estimated from the observed stay events under the core settings and are, therefore, not listed as scalar defaults. Instead, we report their estimation choices (e.g., smoothing strength, time-block definition, mixture/fallback rules), which fully determine the fitted artifacts. The fitted artifacts themselves are released as output files (per scenario folder) and can be reproduced exactly given Table A.1 and the fixed RNG control.

*Randomness control*. All Monte Carlo sampling is controlled by a fixed RNG seed (RANDOM_SEED = 20251209) to ensure exact reproducibility. The seed is an arbitrary integer identifier chosen for bookkeeping (here, a date-like experiment tag) and carries no statistical meaning. For paired scenario comparisons, the RNG is reset per scenario (RESET_SEED_PER_SCENARIO = True) so that differences across scenarios are attributable to POI-inventory changes rather than random draws.

Table A.1. Core settings used in all main experiments.

| Component | Setting (symbol / code name) | Value (main runs) | Purpose / interpretation |
|---|---|---|---|
| Monte Carlo scale | Simulated users per run (SIM_USERS_N) | 5000 | Sample size for Monte Carlo averages |
| Monte Carlo scale | Runs (MC_RUNS) | 50 | Repeated runs for uncertainty reduction |
| Monte Carlo scale | Person-days per user (PERSONDAYS_PER_USER) | 1 | One simulated person-day per simulated user per run |
| Chain safety | Max stay events per person-day (MAX_EVENTS) | 48 | Hard cap to prevent pathological loops |
| Reproducibility | Base RNG seed (RANDOM_SEED) | 20251209 | Arbitrary integer identifier used solely for determinism(no statistical meaning). |
| Reproducibility | Reset seed per scenario (RESET_SEED_PER_SCENARIO) | True | Paired scenario comparisons (same RNG stream per scenario) |
| Start constraint input | Hour×category target matrix ($S_{ipf}$) | Input file | Defines priors $p(h)$ and $p(c|h)$ for the first stay event |
| Start mechanism | Spatialized start POI sampling (USE_SPATIAL_START) | True | Samples the first POI from category-specific POI prior weights (start-stage) |
| Start prior strength | Start prior mixing (START_LAMBDA) | 0.70 | Controls strength of start-stage prior multiplier |
| Start prior shape | Start prior exponent (START_BETA) | 0.70 | Controls concentration/contrast of start-stage prior multiplier |
| Termination | Stop hazard enabled (USE_STOP_HAZARD) | True | Applies hour-dependent termination probability after each stay event |
| Termination | Hazard scale (HAZARD_SCALE) | 1.0 | Multiplies the empirical hazard (1.0 in main runs) |
| Time inhomogeneity | Use time-block transitions (USE_T_BLOCK) | True | Uses block-specific transition kernels $T^{(b)}$ |
| Time inhomogeneity | Time-block edges (BLOCK_EDGES) | e.g.,[0,5,8,11,14,18,24] | Defines $b(h)$ and block-specific $T^{(b)}$ |
| Transitions | Dirichlet smoothing (ALPHA) | 0.5 | Added before row-normalization of expected transition counts |
| Dwell model | Mixture enabled (USE_DWELL_MIXTURE) | True | Fits 2-component GMM on $\log(dwell)$ when evidence is sufficient |
| Dwell model | Mixture components (GMM_COMPONENTS) | 2 | Captures short/long modes |
| Dwell model | Effective evidence threshold (DWELL_MIN_EFFW_HOUR) | 60.0 | Fit hour×category locally only if $W_{eff} \geq$ threshold |
| Dwell model | Shrinkage weight (DWELL_SHRINK) | 0.5 | Convex shrinkage toward category-level/global parameters |
| Dwell model | Minimum dwell (minutes) (MIN_DWELL_MIN) | 5.0 | Lower bound to avoid degenerate durations |
| Time update rule | End-of-day cap | 23:59:59 (same day) | Truncates events crossing day boundary; chain ends at cap |
| Candidate retrieval | Category-specific KNN (POI_K_NEIGH) | 40 | BallTree K-nearest candidates within category |
| Spatial | Exploration probability | 0.02 | Occasional radius-based exploration |
| exploration | (POI_EXPLORE_EPS) | | (≤3 km) to avoid local trapping |
| Spatial exploration | Exploration radius (POI_EXPLORE_RADIUS_M) | 3000m | Radius used when exploration is triggered |
| Likelihood kernel | Distance scale (R_DEFAULT, R_ACCOM) | 100m; 120m | Category-specific $R_c$ in distance-decay likelihood proxy |
| Likelihood kernel | Distance power (POI_DIST_POWER = γ) | 0.75 | Controls tail of decay $\exp(-(d/R_c)^\gamma)$ |
| Likelihood kernel | Uniform mixing (POI_UNIFORM_MIX) | 0.06 | Mixes a small uniform component into distance weights |
| Weak spatial prior | Enable weak POI prior (USE_SOFT_SPATIAL_PRIOR) | True | Multiplies candidate weights by a learned POI "visit prior" (regularizer) |
| Weak spatial prior | Real→POI match radius (PRIOR_MATCH_RADIUS_M) | 80m | Radius for accumulating weak prior evidence from observed stay events |
| Weak spatial prior | Prior mixing (PRIOR_LAMBDA) | 0.60 | Strength of the prior multiplier in candidate weighting |
| Weak spatial prior | Prior exponent (PRIOR_BETA) | 0.60 | Concentration of the prior multiplier |
| Numeric stability | Epsilon stabilizer (PRIOR_EPS = ε) | $1e^{-12}$ | Prevents 0/0 in normalization and stabilizes extreme sparsity |
| Scenario emphasis | Scenario boost (USE_SCENARIO_BOOST) | *True* | Upweights POIs whose IDs are "changed" vs. baseline inventory |
| Scenario emphasis | Boost factor (POI_BOOST_FACTOR) | 3.0 | Multiplier strength applied to changed POIs within candidate sets |
| Zone emphasis | Yumoto zone boost (USE_YUMOTO_ZONE_BOOST) | True | Softly reweights prior for POIs inside Yumoto 3 km zone |
| Zone emphasis | Zone radius (YUMOTO_R_M) | 3000m | Defines the reference anchor zone (3 km) around Hakone-Yumoto. |
| Zone emphasis | Zone mixing (ZONE_LAMBDA) | 0.50 | Strength of zone-factor adjustment |
| Zone emphasis | Zone exponent (ZONE_BETA) | 0.50 | Concentration/contrast of zone-factor adjustment |

Dwell-time fitting (implementation-level). For each hour–category cell $(h, c)$, we use the soft label $p_c(e)$ of each observed stay event $e$ starting at hour $h$ as its contribution weight. The effective weighted evidence is defined as $W_{eff} = \sum_n p_{n,c} \max(0, p_c(e))$. If $W_{eff} \geq$ DWELL_MIN_EFFW_HOUR(60.0), we fit a two-component Gaussian mixture to $\log d$ via resampling proportional to $p_c(e)$ (with GMM_COMPONENTS = 2) and apply convex shrinkage toward the category-level model using DWELL_SHRINK (0.5). Otherwise, we fall back to the corresponding category-level/global model. Simulated dwell times are truncated by MIN_DWELL_MIN (5.0) and the end-of-day cap.

**Figure captions**
Fig. 1. Study area and POI categories in Hakone. The boundary defines the analysis domain for stay-event filtering and POI inventory curation. POIs are labeled by MID10 categories; main transport corridors and Hakone-Yumoto Station are shown for orientation.

Fig. 2. Overview of the rhythm-consistent simulation framework. Train-side artifacts are computed once from observed stay events and reused across baseline and scenario runs, which swap only the POI inventory.

Fig. 3. Conceptual illustration of candidate POI retrieval and posterior-like weighting. A stay-event centroid defines a candidate set within a category; relative weights combine distance-decay likelihood and weak prior information before normalization and sampling.

Fig. 4. Spatial footprint of observed vs. simulated stay events (MID10-colored points) in Hakone under the November 2021 baseline setting.

Fig. 5. Category-wise diurnal profiles of normalized dwell minutes (EDM share) for observed and simulated stay events.

Fig. 6. Day-by-hour heatmaps of normalized EDM (per-category normalized sum) for observed and simulated stay events.

Fig. 7. Day-by-hour heatmaps of normalized ES (per-category normalized sum) for observed and simulated stay events.

Fig. 8. Residual heatmaps for ES (phase-aligned simulation median minus observation), highlighting systematic deviations across categories and time windows.

Fig. 9. Baseline spatial distribution of aggregated dwell minutes (EDM) over the analysis grid.

Fig. 10. Scenario-induced spatial changes in EDM (scenario minus baseline) for S1-S4.

Fig. 11. Baseline spatial distribution of event counts (ES) over the analysis grid.

Fig. 12. Scenario-induced spatial changes in ES (scenario minus baseline) for S1-S4.